\documentclass[prb,twocolumn,showpacs,preprintnumbers,amsmath,amssymb,groupedaddress]{revtex4}
\usepackage{bm}
\usepackage{amsmath}
\usepackage[dvips]{graphicx}
\begin{document}
\bibliographystyle{apsrev}

\title{Electron-Electron Interactions   in the Vacuum Polarization   
 of Graphene}

\author{Valeri N. Kotov}
\affiliation{Department of Physics, Boston University, 590 Commonwealth Avenue, Boston,
 Massachusetts 02215}

\author{Bruno Uchoa}
\affiliation{Department of Physics, Boston University, 590 Commonwealth Avenue, Boston,
 Massachusetts 02215}

\author{A. H. Castro Neto}
\affiliation{Department of Physics, Boston University, 590 Commonwealth Avenue, Boston,
 Massachusetts 02215}

\date{\today}
\begin{abstract}
We discuss the effect of electron-electron interactions   on the static
 polarization properties of graphene  beyond RPA.
Divergent self-energy corrections are naturally absorbed into
 the renormalized  coupling constant $\alpha$.  
 We find that the lowest order vertex correction, which is  the first
 non-trivial correlation contribution,  is finite, and about  $30\%$ 
 of the RPA result  at strong coupling $\alpha \sim 1$.  The vertex  correction 
  leads to further  reduction of the effective charge.
Finite contributions to dielectric screening are expected in all
 orders of perturbation theory. 
\end{abstract}

\maketitle

\section{Introduction and Motivation}

Graphene is a two-dimensional (2D) allotrope of carbon on a honeycomb lattice
with one electron per $\pi$ orbital (half-filling). Its bare electronic
spectrum is described in terms of a linearly dispersing, massless, chiral 
Dirac field ($\Psi_{\bf{p}}$). Since its isolation a few years ago \cite{Graphene} it was realized 
that graphene displays a number of unique properties that are at odds with the standard
theory of metals. \cite{Graphene,Peres,AntonioRMP}
 One of the most important unresolved questions in graphene
is  the role of electron-electron interactions. \cite{Paco1,Paco2,EM,Barlas,Barlas2,NFL1,NFL2,Son,Joerg}
Even though, due to the vanishing of the density of states at the Fermi energy,
the electron-electron interactions are expected to remain unscreened and strong, it is not clear 
what is their influence in the properties of graphene.

In the present work we study the influence of the electron-electron interactions 
on the static dielectric function of graphene at half filling. We perform calculations to
  one order beyond the conventional random phase approximation (RPA)
 vacuum polarization bubble, by including self-energy and vertex corrections 
 into the polarization loop. Our main finding
 is that vertex contributions become important in the coupling regime $\alpha \sim 1$,
 which in turn means that other non-RPA contributions should also be included.  
 We were mainly motivated by the question whether the interactions can significantly
affect the screening properties. This issue is particularly relevant in graphene
 for two reasons: (1.) The effective coupling constant $\alpha$
 (see the precise definition below) in graphene is large $\alpha \sim 1$, and thus
 interactions are expected to be generically important, and
(2.) Despite of the above, to the best of our knowledge, no clear signatures
 of interaction effects have been observed so far in graphene.
 For example measurements of the compressibility \cite{Yacoby} have not detected
 electron correlation effects. In addition, screening  of  external 
 charged impurities introduced in graphene is also expected to
 be sensitive to interaction effects, at least on theoretical
 level, \cite{Impurity1,Impurity11,Impurity2,Impurity3}  and could be relevant for interpretation
of   recent experiments on charged impurity scattering.
 \cite{ImpurityExp}  It is thus
generally important  to investigate the problem of how the correlations
 affect the effective charge of the carriers in graphene, which is determined
by  the vacuum polarization. We will assume that graphene  at half filling
(i.e. when the chemical potential crosses the Dirac point)
 remains a homogeneous gas of quasiparticles, which in itself is 
 not necessarily an innocent assumption due to the possibility of
 puddles, ripples, etc. \cite{AntonioRMP} However we assume that the system
 is homogeneous as the importance of the above effects is still unsettled.

Our starting point is the  low-energy Hamiltonian of graphene  which
can be written as (we use units such that $\hbar=1$),
\begin{equation}
\label{ham1}
H = \sum_{\bf{p}} \Psi_{\bf{p}}^{\dagger} ( v |{\bf p}| \hat{\sigma}_{\bf p} - \mu \hat{\sigma}_{0}) 
\Psi_{\bf{p}}  +  H_{I},
\end{equation}
where $v$ is the Fermi velocity, $\mu$ is the chemical potential away from half-filling, 
$\hat{\sigma}_{0} = \hat{I}$ is the 2$\times$2 identity matrix,
$\hat{\sigma}_{\bf p} \equiv  \hat{\sigma}\cdot{\bf p}/|{\bf p}| =
 (\hat{\sigma}_{x} p_x + \hat{\sigma}_{y} p_y)|{\bf p}|^{-1} $,
 and $\hat{\sigma}_{x},\hat{\sigma}_{y}$ are Pauli matrices.
The first term in the Hamiltonian (\ref{ham1}) reflects the effective Lorentz invariance that exists 
in the non-interacting problem at low energies and gives rise to
bizarre electronic behavior analogous to the one found in quantum electrodynamics (QED). \cite{pw}

In Eq.(\ref{ham1}), $H_{I}$ is the electron-electron interaction,
\begin{equation}
\label{ham2}
H_{I} = \frac{1}{2} \sum_{\bf{p}} \hat{n}_{\bf p} V_{{\bf p}}  \hat{n}_{\bf-p}, \
\  \hat{n}_{\bf p} \equiv \sum_{\bf{q}} \Psi_{\bf{q+p}}^{\dagger} \Psi_{\bf{q}}, 
\end{equation}
 where 
\begin{equation}
\label{Coulomb}
V_{{\bf p}}=\frac{2\pi e^{2}}{|{\bf p}|}
\end{equation}
 is the Fourier transform of the Coulomb potential in 2D. 
 The relative strength of the Coulomb interactions to the kinetic energy
is determined by graphene's ``fine structure constant" $\alpha \equiv e^{2}/v$.
Unlike QED, the Dirac fermion velocity is much smaller than the speed of light, $c$, and hence the Coulomb
field can be treated as instantaneous ($v \approx 10^6 \mbox{m/s}$).
As a result, the Coulomb interaction
breaks the Lorentz invariance of the problem leading to fundamental differences between the graphene problem
and QED. From now on we absorb the dielectric constant of the
 medium $\varepsilon$ into the definition of the effective charge $e$.  For example in the typical case of
  graphene on a SiO$_2$ substrate with dielectric constant $\varepsilon \approx 4$, we have
 $e^{2} = 2e_{0}^{2}/(1+\varepsilon)$, where $e_{0}$ is the charge of the electron.
Keeping in mind that $e_{0}^{2}/v \approx 2.2$, one then finds 
 the coupling constant $\alpha \approx 0.9$. \cite{remark} Nevertheless, even in this situation
 when the relation $\alpha \ll 1$ is not strictly satisfied, perturbation theory is expected
 to give a good
 indication for the behavior of physical quantities. 

The rest of the paper is organized as follows. Section II deals
with corrections to the polarization loop arising from the dressing of the electron
 propagators. In the Section III the true interaction (correlation) insertion, the vertex correction,
 is examined. Section IV contains our conclusions.

\section{Self-energy corrections to the polarization}

We concentrate on the most interesting case of zero Fermi energy
($\mu \rightarrow 0$), when the low-energy physics is controlled by the 
proximity to the Dirac point. 

The free Dirac Green's function is 
\begin{equation}
\label{gf1}
\hat{G}({\bf k},\omega) = \frac{1}{ \omega \hat{\sigma}_{0} - v|{\bf k}|\hat{\sigma}_{\bf k} + 
 i \hat{\sigma}_{0} 0^{+} {\mbox{sign}} (\omega)}.
\end{equation}
The interaction effects lead to the dressed Green's function
$\hat{G}^{-1} \rightarrow \hat{G}^{-1} -  \hat{\Sigma} $, where the self-energy  $\hat{\Sigma}$ 
 is a sum of two terms with different matrix structure:
$\hat{\Sigma} = \hat{\Sigma}_{0} +{\bf{\hat{\Sigma}}}$,
$\hat{\Sigma}_{0} \propto \hat{\sigma}_{0}, {\bf{\hat{\Sigma}}} \propto  \hat{\sigma}_{\bf k}$.  
 At Hartree-Fock (HF) level (first order in $\alpha$) a divergent contribution appears,
 due to the long-range nature of the Coulomb interaction \cite{Paco1}
where $\Lambda \sim 1/a \gg k$ is an ultraviolet cutoff ($a$ is the lattice
 spacing). One finds
\begin{equation}
\label{hf}
 {\bf{\hat{\Sigma}}}^{(1)}(|{\bf k}|) = (\alpha/4) \ (v|{\bf k}|)
 \hat{\sigma}_{\bf k}  \ln(\Lambda/|{\bf k}|),
\end{equation}
 which implies that the effective velocity  changes 
$v \rightarrow v(1 + (\alpha/4) \ln(\Lambda/|{\bf k}|))$, and grows without bound
 at low energies $|{\bf k}|/\Lambda \rightarrow 0$. 
 This should in principle lead to  anomalies in thermodynamic and spectral properties 
of graphene. \cite{Paco1,vafek}  From theoretical viewpoint,  most importantly, the single
 logarithmic behavior was found to persist to second order of perturbation
 theory as well, \cite{EM} and consequently this is expected to be the case to all orders,
 reflecting the fairly simple (at least at weak-coupling) renormalization structure of the theory. 

 We now turn to the calculation of the static polarization, $\Pi({\bf q})\equiv \Pi({\bf q},\omega=0)$.
The frequency variable in $\Pi({\bf q})$ is omitted from now on. 
The bare polarization bubble (without any interaction lines in the loop)  is known to be 
\begin{eqnarray}
\Pi^{(0)}({\bf q}) &=& -i \sum_{\bf{k}} \int \frac{d\omega}{2\pi} {\mbox{Tr}} \{ \hat{G}({\bf k},\omega)
\hat{G}({\bf k + q},\omega)\} \nonumber \\
&& = - |{\bf q}|/(4v).
\end{eqnarray}
From now on  the trace   stands for summation  over spin (s), valley (v) and pseudospin (Pauli matrix
$\sigma$) indices, i.e.
\begin{equation}
\label{trace}
{\mbox{Tr}} = \sum_{s,v} {\mbox{Tr}_{\sigma}} = 4 {\mbox{Tr}_{\sigma}}.
\end{equation}
The momentum sums are performed as  $\sum_{\bf{k}}=\int d^{2}k/(2\pi)^{2}$.

\begin{figure}[bt]
\centering
\includegraphics[height=88pt, keepaspectratio=true]{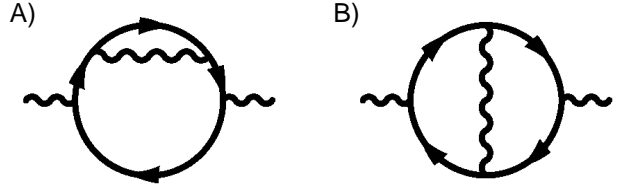}
\caption{First order interaction corrections to the polarization bubble: (a.) self-energy
 correction, (b.) vertex correction. The explicit form of these diagrams is given by
equations (9) and (19). 
 The wavy line represents the Coulomb interaction, Eq.~(3).}
\label{Fig1}
\end{figure}

Next, we calculate the bubble dressing due to the electron-electron interactions to
 first order in $\alpha$. The two diagrams at this order are shown in Fig.~\ref{Fig1}.
We write the total polarization as
\begin{equation}
\Pi({\bf q}) = \Pi^{(0)}({\bf q}) + \Pi^{(1)}({\bf q}) + \Pi^{(2)}({\bf q}),
\end{equation}
where  $\Pi^{(1)}$ and $\Pi^{(2)}$ stand for the contributions of Fig.~\ref{Fig1}(a)  and Fig.~\ref{Fig1}(b),
 respectively. 
\noindent

The self-energy dressing of Fig.~\ref{Fig1}(a) can be written  as
(the factor of $2$ originates from the two possible insertions)
\begin{eqnarray}
\label{pol1}
&\Pi^{(1)}({\bf q}) &= -2i \sum_{\bf{k}} \int \frac{d\omega}{2\pi} {\mbox{Tr}} \{ \hat{G}({\bf k},\omega)
\hat{G}({\bf k + q},\omega)\hat{G}({\bf k},\omega) \nonumber \\
&& \times \hat{\Sigma}({\bf k},\omega) \}.
\end{eqnarray}
At lowest order, the self-energy is simply the Hartree-Fock one, meaning that in (\ref{pol1})
 we replace 
\begin{equation}
\label{se}
\hat{\Sigma}({\bf k},\omega) \rightarrow
 \hat{\Sigma}^{(1)}({\bf k}) = i \sum_{{\bf p}} \int  \frac{d\omega_{1}}{2\pi} \hat{G}({\bf p},\omega_{1})
 V_{\bf{k}-\bf{p}}. 
\end{equation}
 The large logarithm present in $\hat{\Sigma}^{(1)}({\bf k})$ at low momenta,  Eq.~(\ref{hf}), is expected to
 appear also in some form in  $\Pi^{(1)}({\bf q})$.

 Let us define the following quantities which appear in our results from now on
\begin{equation}
\label{notation}
\Delta(\bf{k},\bf{q})  \equiv  1 - \frac{\hat{\bf{k}}\cdot (\bf{k}+ \bf{q})}{ |\bf{k}+ \bf{q}|},
\ \ \ \hat{{\bf k}} \equiv {\bf k}/|{\bf k}|,
\end{equation}  
\begin{equation}
E({\bf{k}},{\bf{q}})  \equiv v(|{\bf{k}}| + |\bf{k}+ \bf{q}|).
\end{equation}
After performing the frequency, and then momentum integrations in (\ref{pol1}),
and using the self-energy from (\ref{se}), we obtain
\begin{eqnarray}
\label{pol10}
&\Pi^{(1)}({\bf q}) &=  4 \sum_{\bf{k},\bf{p}} V_{\bf{k}-\bf{p}} (\hat{\bf{k}} \cdot \hat{\bf{p}})
 \frac{\Delta({\bf k},{\bf q})}{[E({\bf k},{\bf q})]^{2}}
 \nonumber \\
&&  = \frac{\alpha}{16} \ \frac{|{\bf q}|}{v} \ \ln(\Lambda/|{\bf q}|), \ \ \Lambda/|{\bf q}| \gg 1.
\end{eqnarray}
This result means that  the large logarithm in $\Pi^{(1)}({\bf q})$ simply reflects the
 renormalization of the Fermi velocity, i.e. this divergence is not independent,
 but can be simply reabsorbed into the velocity by replacing
$v \rightarrow v(1 + (\alpha/4) \ln(\Lambda/|{\bf q}|))$ in the one-loop result
 $\Pi^{(0)}({\bf q}) = - |{\bf q}|/(4v)$. Due to the simple logarithmic structure of
 the theory this is expected to hold to all orders of perturbation theory,
 i.e. all self-energy corrections lead to a replacement of the coupling
 $\alpha$ in all final expressions with the ``running" coupling  $\alpha(q)$,
 accounting for the velocity renormalization.
 We  therefore  assume that the velocity renormalization procedure 
is performed in all higher order diagrams.
At finite   small  chemical potential, $\mu \ll \Lambda$,  which is the case in
 any realistic experimental situation, the divergence is cut-off, $\ln(\Lambda/|{\bf q}|)
\rightarrow \ln(\Lambda/\mu)$. Due to the slow variation of the logarithmic function
 or possible other factors (such as strong dielectric screening),   no significant
 variation of the velocity has been found in experiment. \cite{AntonioRMP}

An interesting effect, related to the  interaction contribution $\Pi^{(1)}({\bf q})$,
Eq.~(\ref{pol10}), was recently
 discussed in Ref.~\onlinecite{Impurity2} within the RG approach. Our calculation,
 leading to Eq.~(\ref{pol10}), provides an explicit perturbative confirmation of the RG 
results. 
 If we imagine an external Coulomb impurity with charge $Z>0$, probing the
 polarization of the vacuum, then the induced  charge density, in momentum space,   is
$\rho_{ind}({\bf q}) = Z V_{{\bf q}}\Pi({\bf q})$. Here $\Pi({\bf q}) = \Pi^{(0)}({\bf q}) +
 \Pi^{(1)}({\bf q})$. While the first term leads to induced charge $\rho_{ind}^{(0)}({\bf q})
= -Z\alpha (\pi/2)$, localized in real space at the impurity site and with a screening sign, the
 interaction term  $\rho_{ind}^{(1)}({\bf q}) = Z\alpha^{2} (\pi/8) \ln(\Lambda/|{\bf q}|)$
 has an opposite sign and decays as an inverse power law ($1/r^{2}$, with logarithmic corrections). 
 This peculiar behavior simply reflects, however,
the renormalization (increase) of the Fermi velocity at low momenta, which
 leads to suppression of screening at large distances.
\begin{figure}[bt]
\centering
\includegraphics[height=115pt, keepaspectratio=true]{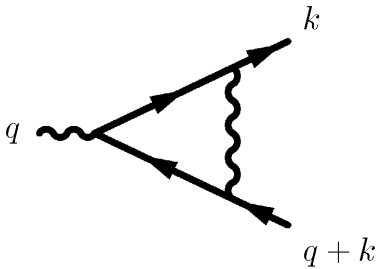}
\caption{First order interaction correction to the vertex $\hat{\Gamma}(k;q)$,
 given by Eq.~(\ref{vertex}).}
\label{Fig2}
\end{figure}
\noindent

\section{Vertex corrections to the polarization bubble}

 We proceed with  the calculation of the vertex correction in  Fig.~\ref{Fig1}(b).
 Before evaluating this expression, it is useful to examine the
 (possible) singularity structure separately in the vertex operator,
 shown in Fig.~\ref{Fig2}. For simplicity  we use the notation
$\hat{\Gamma}(k;q)$ which stands for the more conventionally used
 form $\hat{\Gamma}(k,k+q;q)$, where the variables denote both frequency and
 momenta ($q$ is the bosonic momentum/frequency).

 Since the Coulomb interaction is non-retarded, we have the simple expression 
\begin{equation}
\label{vertex}
\hat{\Gamma}({\bf k};{\bf q}, \omega) =  i \sum_{{\bf p}} \int  \frac{d\omega_{1}}{2\pi}
\hat{G}({\bf p},\omega_{1}) \hat{G}({\bf p}+{\bf q},\omega_{1}+\omega)  V_{\bf{k}-\bf{p}}.
\end{equation}
After evaluating the frequency integral, the result is a sum of an off-diagonal
 and diagonal parts (with respect to the Pauli matrix indexes),
 $\hat{\Gamma}=\hat{\Gamma}^{(o)} + \hat{\Gamma}^{(d)}$. More explicitly,
\begin{eqnarray}
\label{vertex1}
&\hat{\Gamma}^{(o)}({\bf k};{\bf q}, \omega)& \propto \sum_{{\bf p}}  V_{\bf{k}-\bf{p}}
(\hat{\sigma}_{\bf p} - \hat{\sigma}_{\bf p + q}) \left ( \frac{1}{E(\bf{p},\bf{q}) + \omega}
\right. \nonumber \\
&& \left.
- \frac{1}{E(\bf{p},\bf{q}) - \omega} \right ).
\end{eqnarray}
We are interested only in the zero frequency limit (and only in the real part of $\hat{\Gamma}$,
since the imaginary part does not contribute to the polarization). In this case 
 the off-diagonal piece vanishes identically, $\hat{\Gamma}^{(o)}({\bf k};{\bf q}, \omega=0)=0$.
This is expected to be the case since the Coulomb interaction is diagonal and thus the vertex
 can not generate a static contribution with a different matrix structure.
On the other hand the diagonal part is finite in the same limit
\begin{equation}
\label{vertex2}
\hat{\Gamma}^{(d)}({\bf k};{\bf q}, \omega=0) \propto  \sum_{{\bf p}}  V_{\bf{k}-\bf{p}}
(1-\hat{\sigma}_{\bf p}\hat{\sigma}_{\bf p + q}) \frac{1}{E(\bf{p},\bf{q})}.
\end{equation}
 An explicit evaluation shows that $\hat{\Gamma}^{(d)}({\bf k};{\bf q}, \omega=0)$ does not
 have any divergent contributions. For example at
$|{\bf k}| \ll |{\bf q}|$,
\begin{equation}
\hat{\Gamma}^{(d)}({\bf k};{\bf q},0) \propto \alpha \ {\mbox{const.}}, \ \ |{\bf k}| \ll |{\bf q}|,
\end{equation}
 while in the opposite limit
\begin{equation}
\hat{\Gamma}^{(d)}({\bf k};{\bf q},0) \propto \alpha (|{\bf q}|/|{\bf k}|), \ \ |{\bf q}| \ll |{\bf k}|
\end{equation}
For our purposes  the exact formulas are not important  (we also do not show the dependence on the
 angle between ${\bf k},{\bf q}$); our main conclusion at this stage is that
 the vertex does not have any divergent parts. We have also examined diagrams of  
 higher order, such as ``ladder" and ``crossed" ladder vertex corrections, and have found that
 all of them are finite. 
 Therefore the  vertex insertions into the polarization function are expected to give a finite 
 contribution to that quantity, and below  we  evaluate the lowest order vertex correction  numerically.
 
 It is clear that a Ward identity relating divergent
 contributions in the self-energy and in the vertex does not hold
 here, unlike   conventional QED  where Lorentz (and gauge) invariance
guarantees cancellation between vertex and self-energy corrections, \cite{IZ}
and charge is renormalized only through simple polarization loops in the photon
 propagator. 
 On the other hand in graphene, where the only non-trivially renormalized quantity is
 the velocity $v$, both the polarization operator and the vertex operator do
 not show any independent divergencies.
 
The diagram of Fig.~\ref{Fig1}(b) now reads  
\begin{equation}
\label{pol2}
\Pi^{(2)}({\bf q})  =  -i \sum_{\bf{k}} \! \int \! \frac{d\omega}{2\pi} {\mbox{Tr}} \{
 \hat{G}({\bf k},\omega)\hat{\Gamma}({\bf k};{\bf q},0)  \hat{G}({\bf k + q},\omega)
\},
\end{equation}
where the full expression for $\hat{\Gamma}$ from Eq.~(\ref{vertex}) should be used.
An explicit calculation, starting by evaluation of the energy integrations, leads to the  result
\begin{equation}
\Pi^{(2)}({\bf q}) = - \frac{1}{4} {\mbox{Tr}}  \sum_{{\bf k},{\bf p}} V_{{\bf k}-{\bf p}}   
\frac{(1-\hat{\sigma}_{\bf p +q}\hat{\sigma}_{\bf p}) (1-\hat{\sigma}_{\bf k}\hat{\sigma}_{\bf
 k + q})}{E({\bf k},{\bf q}) E({\bf p},{\bf q})}.
\end{equation}
Taking into account:
\begin{equation}
\hat{\sigma}_{\bf k}\hat{\sigma}_{\bf p} = \frac{1}{|{\bf k}||{\bf p}|} \left ({\bf k}.{\bf p} +
i\hat{\sigma}_{3} ({\bf k}\times {\bf p})_z \right ),
\end{equation}
where $({\bf p}\times {\bf q})$ stands for a vector product, we then arrive at the final formula
\begin{eqnarray}
\label{pol20}
&\Pi^{(2)}({\bf q})& = -2 \sum_{{\bf k},{\bf p}} \frac{V_{{\bf k}-{\bf p}}}{E({\bf k},{\bf q}) E({\bf p},{\bf q})}
 \left \{ \Delta({\bf k},{\bf q})\Delta({\bf p},{\bf q}) \right. \nonumber \\
&& \left. + \ \frac{({\bf p}\times {\bf q})_{z}({\bf k}\times {\bf q})_{z}}{|{\bf p}||{\bf k}||{\bf p}+
{\bf q}||{\bf k}+{\bf q}|} \right \}.
\end{eqnarray}
It is clear on dimensional grounds that $\Pi^{(2)}({\bf q})$ is linear in $|{\bf q}|$.  This  is in fact 
the case for polarization diagrams in all orders of perturbation theory. 
The four-dimensional integrals, appearing in (\ref{pol20}), cannot be evaluated analytically.
We have found, as expected in light of our previous discussion of the vertex function,
 that the expressions converge in the ultraviolet limit. 
After computing the integrals numerically, we obtain the following result for
 the combination $V_{{\bf q}}\Pi^{(2)}({\bf q})$, which appears in the dielectric function, 
\begin{equation}
\frac{2\pi e^{2}}{|{\bf q}|}\Pi^{(2)}({\bf q}) = - 0.53 \alpha^{2} .
\end{equation}
 Adding also the one-loop RPA result, we have finally (where ${\cal{E}}$ is the static
 dielectric constant, defined by the formula below)
\begin{equation}
\label{charge}
V_{{\bf{q}}}^{{\mbox{eff}}}  =   \frac{V_{{\bf{q}}}}{1- V_{{\bf{q}}}\Pi({\bf q})} = \frac{V_{{\bf{q}}}}{{\cal{E}}},
\end{equation}
\begin{equation}
\label{diel}
  {\cal{E}} = 1 + \frac{\pi}{2} \alpha + 0.53 \ \alpha^{2} +O(\alpha^{3}).  
\end{equation}

We conclude that, at $\alpha \sim 1$, the vertex correction is more than $30\%$ of the
 one-loop result. It also has a screening sign, i.e. it reduces
 the effective charge. One also expects that finite contributions will appear
 to all orders in $\alpha$. However, resummation of perturbation theory
 by simple means seems impossible, as the contributions in question
 are finite and accumulate over a wide range of momenta in the corresponding
 diagrams (rather than within a specific integration window, from where divergent parts
 typically originate, and thus can be easily collected). 
Even though the vertex contribution is a sizable one, two remarks are in order:
 (1.) It does not change drastically the structure of the theory, apart
 from contributing towards further screening of the interactions. 
 (2.) The fact that perturbation theory is used with the intention of being
 applied at a rather strong coupling is in itself questionable.
 Nevertheless, perturbation theory provides a clear indication that a
 significant contribution to screening exists beyond the conventional 
 one-loop RPA result. On the other hand in the weak-coupling regime, $\alpha \ll 1$,
 RPA is  parametrically well  justified as far as the static polarization properties
 are concerned (although   the RPA is not justified  for the self-energy. \cite{EM})  

\section{Discussion and Conclusions}

  It is also useful to compare our results to the situation in ordinary metals
 with a finite Fermi surface. Certain approximations are typically used to
 account for vertex corrections, such as the Hubbard form of the dielectric
 function. When extrapolated to low momentum, the vertex contribution
 tends to  decrease the screening length,   \cite{Mahan} i.e. it reduces further the range
 of the interactions.  Naturally in graphene, where  the screening length is
 infinite (for the case of zero chemical potential considered here),
 the vertex correction affects directly the effective charge, without
changing the shape of the Coulomb potential.

Finally we mention two recent related works, discussing interaction effects, 
that appeared while the present manuscript was being prepared.
 In Ref.~\onlinecite{Mi2}, the effect of self-energy and vertex corrections
to lowest order ($\alpha$) on the minimal conductivity in graphene was discussed,
 with the conclusion that the corrections is of order $1\%$.
 Dynamical polarization properties  were studied in Ref.~\onlinecite{Mi3},
 where the vertex diagrams were found to have logarithmically singular
 contributions near the threshold $\omega \sim vq$, leading to the possibility
 of a plasmon mode.

In summary, we have shown that vertex corrections can have sizable effect
in the static vacuum polarization diagrams in the regime of strong coupling, 
while for small coupling their importance diminishes parametrically.
The  self-energy corrections are naturally absorbed into
 the renormalization of the Fermi velocity. 
 The non-RPA vertex diagram at lowest order of perturbation theory 
 was found to  decrease the effective charge, meaning that in principle
 correlation effects at higher order must  also be taken into account. 
Thus the ultimate asymptotic behavior of the static polarization function  for $\alpha \sim 1$
 remains an open problem.

\begin{acknowledgments}
We are grateful to D. K. Campbell,  R. Shankar, V. Pereira, O. Sushkov, and A. Polkovnikov
 for many insightful discussions. B.U.
acknowledges CNPq, Brazil, for support under grant No.201007/2005-3.
\end{acknowledgments}


\begin{thebibliography}{99}
\bibitem{remark} Other substrates can lead to a smaller coupling, e.g. for  SiC,
 $\varepsilon \approx 10$, and $\alpha \approx 0.4$.
\end{thebibliography}
\end{document}